\documentclass[iop]{emulateapj}
\usepackage{amsmath}

\newcommand{\vcyg}{\object{V404 Cygni}}

\shorttitle{GMRT monitoring of V404 Cygni during the June 2015 outburst}
\shortauthors{Chandra \& Kanekar}

\begin{document}

\title{Giant Metrewave Radio Telescope monitoring of the black hole X-ray binary, V404 Cygni, during its June 2015 outburst}

\author{Poonam Chandra\altaffilmark{1} and Nissim Kanekar\altaffilmark{1}}
\affil{National Centre for Radio Astrophysics, TIFR, Pune University Campus, Pune 411007, India}

\begin{abstract}

We report results from a Giant Metrewave Radio Telescope (GMRT) monitoring campaign on 
the black hole X-ray binary V404~Cygni during its 2015 June outburst. The GMRT observations
were carried out at observing frequencies of 1280, 610, 325 and 235~MHz, and extended from 
June~26.89~UT (a day after the strongest radio/X-ray outburst) to July~12.93~UT. We find the 
low-frequency radio emission of V404~Cygni to be extremely bright and fast-decaying in the 
outburst phase, with an inverted spectrum below 1.5~GHz and an intermediate X-ray state. The 
radio emission settles to a weak, quiescent state $\approx 11$~days after the outburst, with a 
flat radio spectrum and a soft X-ray state. Combining the GMRT measurements with flux density 
estimates from the literature, we identify a spectral turnover in the radio spectrum at $\approx 1.5$~GHz 
on $\approx$~June~26.9~UT, indicating the presence of a synchrotron self-absorbed emitting 
region. We use 
the measured flux density at the turnover frequency with the assumption of equipartition of energy 
between the particles and the magnetic field to infer the jet radius ($\approx 4.0 \times 10^{13}$~cm), 
magnetic field ($\approx 0.5$~G), minimum total energy ($\approx 7 
\times 10^{39}$~ergs) and transient jet power ($\approx 8 \times 10^{34}$~erg~s$^{-1}$). 
The relatively low value of the jet power, despite V404 Cygni's high black hole spin parameter, 
suggests that the radio jet power does not correlate with the spin parameter.
 
\end{abstract}


\section{Introduction}

Black hole X-ray binaries (BHXBs) are systems in which a black hole accretes matter from a 
low mass (${\rm M} \approx {\rm M}_{\odot }$) Roche lobe-filling companion \citep[e.g.][]{remillard06}.
BHXBs have very different X-ray properties, in terms of luminosity and spectral shape, depending
on their state. Typical BHXBs spend most of their time in a quiescent state, with a low 
X-ray luminosity $\approx 10^{30.5} - 10^{33.5}$~erg~s$^{-1}$, i.e. $\lesssim 10^{-5} L_{\rm Edd}$, 
and a non-thermal, ``hard'' X-ray spectrum. They are believed to go into an outburst state 
when the accretion rate increases by a few orders of magnitude, e.g., when an instability is 
triggered in the accretion disk \citep[e.g.][]{remillard06,fender14}. The outburst states 
contain both a thermal and a non-thermal X-ray component, and are usually classified as 
``hard'' (low luminosity, dominated by the non-thermal component), ``soft'' (high luminosity, 
dominated by the thermal component), and ``very high'' or ``steep power law'' (high luminosity, 
containing both thermal and non-thermal components, and with a steep photon index, $\approx 2.5$) 
\citep[e.g.][]{remillard06,belloni16}.

Radio emission from BHXBs is believed to arise from jets, and to be closely connected to the black 
hole accretion properties \citep[e.g.][]{fender14}. Weak, flat-spectrum radio emission is sometimes 
seen in the low-accretion, quiescent state of a BHXB, probably arising from a steady compact jet. 
As the accretion rate increases, and a BHXB moves from quiescence to the low-hard state, the 
radio emission steadily increases while retaining its flat spectrum \citep{fender06}. In the 
low-hard state, a remarkable correlation between radio and X-ray luminosities has been seen over 
more than three orders of magnitude in X-ray luminosity \citep[e.g.][]{corbel08,corbel13},
although recent observations suggest the possibility of two separate tracks in this correlation 
\citep{fender14}
This can be naturally explained in a ``jet-accretion coupling'' scenario where the X-ray emission 
arises from a hot accretion flow, while the radio emission stems from relativistic electrons in a jet 
\citep{fender06}. The spectrum remains flat at GHz frequencies in the low-hard state, probably due 
to self-absorption by moderately relativistic electrons at different radii \citep[e.g.][]{foster96}.

When the X-ray luminosity crosses $\sim 10^{37}$~erg~s$^{-1}$, BHXBs typically transition
from a low-hard state to a high-soft state, dominated by thermal X-ray emission. This 
transition phase is marked by the brightest radio flares \citep[e.g.][]{fender14}. This has been 
explained as arising due to a transition from a state with a steady radio jet to one with no jet, 
possibly with a transient increase in the jet Lorentz factor that gives rise to shocks within the 
flow, and hence, to increased radio emission \citep{fender04}. In such outbursts, discrete expanding 
radio-emitting regions have been observed on either side of the central BHXB \citep[e.g.][]{mirabel94,tingay95}.
The radio spectrum at frequencies above a few GHz evolves from optically-thick to optically-thin, 
perhaps due to an expanding emission region \citep[][]{vanderlaan66}. Finally, when the BHXB enters 
the high-soft state, the radio emission significantly reduces, suggesting that the radio jet is 
no longer active.

\tabletypesize{\scriptsize}
\begin{deluxetable*}{lcccccccc}[t!]
\centering
\tablecaption{Details of the GMRT observations of V404~Cygni
\label{tab:obs}}
\tablewidth{0pt}
\tablehead{
\colhead{Mean date of} & \colhead{MJD} & \colhead{Days since}   &\colhead{Central} & \colhead{No. of} & \colhead{On-source} & \colhead{Resolution} & \colhead{RMS noise} & \colhead{Flux density\tablenotemark{a}} 
 \\
\colhead{obsn. (UT)} & & \colhead{outburst\tablenotemark{b}} & \colhead{frequency (MHz)} &  \colhead{antennas\tablenotemark{c}} &  \colhead{time (m)} & \colhead{($''\times''$)}& \colhead{mJy} & \colhead{mJy}
}
\startdata
2015 June 26.89 & 57199.89 & 11.12 & 1280  & 14 & 90   & $2.4 \times 2.0$    & $0.25$ &  $739 \pm 77   $  \\
2015 June 26.89 &  57199.89 &11.12 & 610  & 12 & 90   & $13.0 \times 6.8$   & $0.51$ &  $470 \pm 49   $  \\
2015 June 26.89 & 57199.89 &11.12 & 235  & 12 & 90   & $13.2 \times 10.7$  & $2.6$  &  $188 \pm 27   $  \\
2015 June 27.96 & 57200.96 &12.19 & 325  & 28 & 120  & $10.5 \times 9.6$   & $0.42$ &  $232 \pm 23   $  \\
2015 July 01.93 & 57204.93 & 16.16 & 1280 & 10 & 186  & $3.8 \times 3.3$    & $0.06$ &  $6.39 \pm 0.67$  \\
2015 July 01.93 & 57204.93 & 16.16 & 610  & 15 & 180  & $6.1 \times 5.2$    & $0.12$ &  $8.88 \pm 0.94$  \\
2015 July 01.93 & 57204.93 & 16.16 & 235  & 13 & 180  & $12.6 \times 10.6$  & $1.5$  &  $13.4 \pm 2.4 $  \\
2015 July 07.04 & 57210.04 & 21.27 & 1280 & 13 & 55   & $ 3.7 \times 2.7$   & $0.07$ &  $0.78 \pm 0.13$  \\
2015 July 07.04 &57210.04 & 21.27 & 610  & 14 & 54   & $ 12.0 \times 7.0$  & $0.25$ &  $<0.75        $  \\
2015 July 07.04 &57210.04 & 21.27 & 235  & 11 & 54   & $ 26.0 \times 13.6$ & $3.7$  &  $<11.2        $  \\
2015 July 11.02 & 57214.02 & 25.25 & 1280 & 13 & 96   & $ 4.6 \times 3.1$   & $0.09$ &  $0.52 \pm 0.14$  \\
2015 July 11.02 & 57214.02 & 25.25 & 610  & 12 & 97   & $ 22.7 \times 7.1$  & $0.20$ &  $ <0.60       $  \\
2015 July 12.65 & 57215.65 & 26.88 & 610  & 26 & 68   & $ 9.7 \times 5.2$   & $0.10$ &  $0.66 \pm 0.24$  \\
2015 July 12.65 & 57215.65 & 26.88 & 235  & 24 & 38   & $ 21.1 \times 11.5$ & $1.6$  &  $ <4.7        $  \\
2015 July 12.93 & 57215.93 & 27.16 & 1280 & 13 & 180  & $ 2.9 \times 2.7$   & $0.07$ &  $0.52 \pm 0.12$  \\
2015 July 12.93 & 57215.93 & 27.16 & 610  & 14 & 180  & $ 7.4 \times 5.8$   & $0.12$ &  $0.50 \pm 0.19$  \\
\enddata
\tablenotetext{a}{The quoted errors include measurement errors, and uncertainties in the 
flux density scale and in the ratio of the system temperatures on \vcyg\ and the flux 
calibrators. The upper limits are at $3\sigma$ significance.}
\tablenotetext{b}{Days since the outburst on 2015 June 15.77 UT (MJD 57188.77).}
\tablenotetext{c}{The number of working GMRT antennas at the observing frequency.}
\end{deluxetable*}

A critical ingredient in estimating the total jet power in an outburst is the ``break frequency'' 
at which the jet transitions from being optically-thick to optically-thin. Unfortunately,
few BHXB outbursts have simultaneous monitoring over a sufficiently wide range of radio 
frequencies close to an outburst peak to accurately measure the break frequency. This is 
especially difficult at high frequencies, where the transition to optically-thin behaviour 
arises at very early times, soon after the outburst. Most estimates of jet power are hence
based on single-frequency radio observations, with an assumed low frequency spectral shape; this 
could imply systematic errors in the jet power estimates. 

\vcyg\ is a low-mass BHXB consisting of a black hole of mass $9.0\; {\rm M}_\odot$, 
accreting material from a low-mass ($< 1 \; {\rm M}_\odot$), late-type companion star 
\citep{khargharia10}. It has the longest orbital period of all known BHXBs. 
Due to its proximity \citep[distance~$\approx 2.39$~kpc; ][]{miller-jones09} and high luminosity, 
\vcyg\ is an excellent test-bed to study BHXB accretion and is hence regularly monitored for variability. 
The X-ray monitoring of \vcyg\ revealed a major outburst on 2015 June~15.77~UT 
\citep[e.g.][]{barthelmy15,negoro15,kuulkers15}. This led to extensive follow-up monitoring at 
all available wavebands, and the discovery of a radio outburst ten days after the initial 
X-ray transient. Further, a search of archival data with the 2-m Faulkes Telescope North revealed 
an optical precursor one week prior to the X-ray outburst \citep{bernardini16}. An optical spectrum 
obtained 13~hours prior to the {\it Swift}-BAT outburst found spectral lines typical of an accretion disk.

In this article, we present simultaneous Giant Metrewave Radio Telescope (GMRT) multi-frequency monitoring 
observations of \vcyg\ during the June~2015 event, which allow 
us to estimate the break frequency within a day of the strongest radio outburst, and hence 
to estimate the transient jet power. In \S \ref{obs}, we detail the observations and data analysis. 
Our results and interpretations are discussed in \S \ref{results}.

\begin{figure}[t!]
\includegraphics*[width=0.45\textwidth]{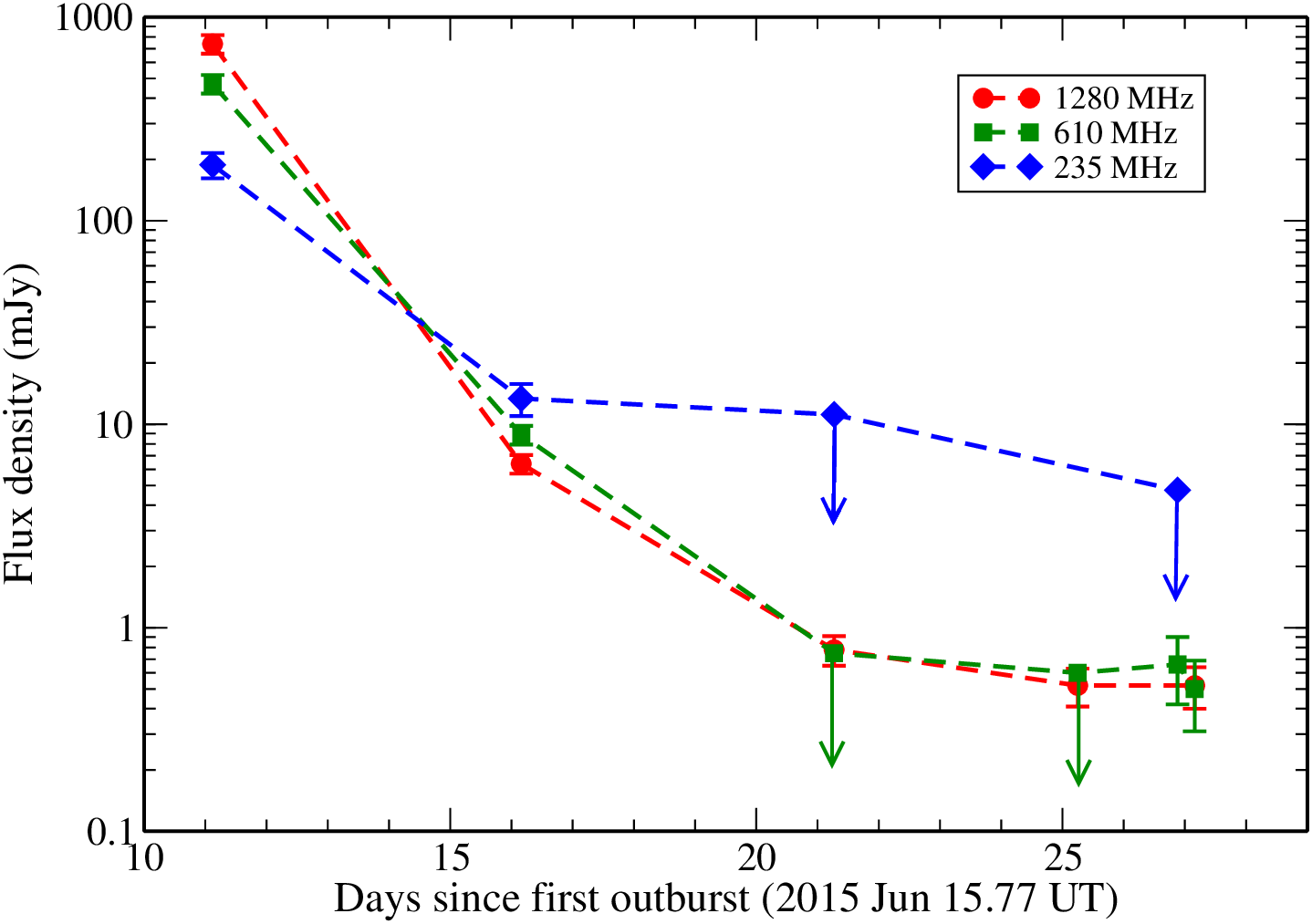}
\caption{The GMRT light curves of V404~Cygni at 1280 MHz (red circles), 610 MHz (green squares) and 
235 MHz (blue diamonds). The downward-pointing arrows indicate $3\sigma$ upper limits on the 
source flux density, in cases of non-detections.
\label{fig:lc}}
\end{figure}

\section{Observations and data analysis}
\label{obs}

Our GMRT monitoring of \vcyg\ began on 2015 June 26.89~UT, with the GMRT antennas split into 
two sub-arrays, consisting of sixteen antennas tuned to an observing frequency 
of 1280~MHz (with two polarizations) and the remaining fourteen antennas tuned 
to the dual 610/235 mode, with observing frequencies of 610~MHz and 235~MHz in the right 
and left circular polarizations, respectively. This setup was repeated for the observations of 
2015~July~01.93~UT and July~07.04~UT. We also observed \vcyg\ on June~27.96~UT at 325~MHz, using 
the full polar mode, and on July~12.65~UT with the above dual 610/235 mode.  Finally, 
our observations on  July~11.02~UT and July~12.93~UT used two sub-arrays, with 16 antennas 
tuned to 1280~MHz and 14~antennas tuned to 610~MHz, both with two polarizations.

All observations used the GMRT Software Backend as the correlator, with a bandwidth of 33.3~MHz 
divided into 256 channels, and with 2~second integrations. Observations of the standard 
calibrators 3C48 or 3C295 at the start and/or the end of the run were used to calibrate the 
flux density scale and the antenna passband shapes. The compact sources J1924+3329 and 
J2015+3710 were used as phase calibrators, and observed for 6~minutes before and after each
30-minute scan on the target source.

The initial data editing and calibration used the {\sc flagcal} software pipeline 
developed for automatic flagging and calibration of GMRT data \citep{prasad12}. 
Following this, the analysis used standard procedures for widefield imaging and 
self-calibration in ``classic'' AIPS. The flux densities of \vcyg\ at the different observing 
frequencies were estimated via a single-Gaussian fit to a small region centred on the source 
in each image, using the AIPS task {\sc jmfit}. 

\begin{figure}[t!]
\includegraphics*[width=0.45\textwidth]{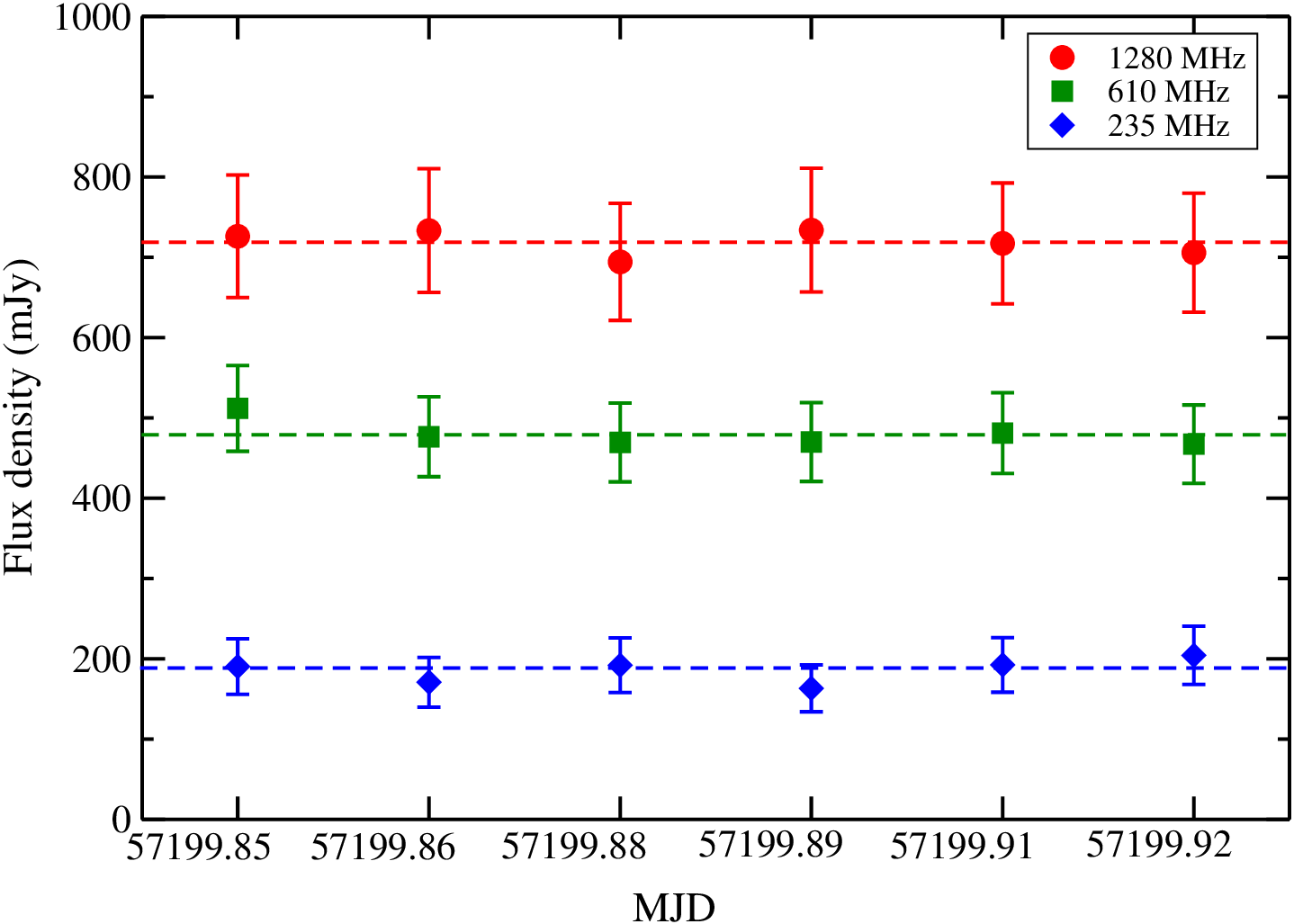}
\caption{The measured flux density of V404~Cygni on 2015 June 26 UT, at 1280, 610 and 235~MHz, in 15-minute bins, 
plotted versus time. Here, the dashed lines are the average flux densities  at the given frequencies
from the full observing run. The figure shows no evidence for statistically-significant intra-epoch variability.
\label{fig:variability}}
\end{figure}

The original GMRT observations used 
automatic level controllers (ALCs) to keep the input power levels of the correlator constant. 
One has to then correct for the fact that antenna gains are inversely proportional to the system 
temperature, which may be different for different sources. The correction factors were estimated 
in April and May 2017 by measuring the ratio of system temperatures on \vcyg\ and the flux 
calibrators with the ALCs switched off. This yielded correction factors of $0.993 \pm 0.032$
(1280~MHz), $1.056 \pm 0.030$ (610~MHz), $1.486 \pm 0.079$ (325~MHz) and $1.36 \pm 0.14$
(235~MHz). The measured flux densities from {\sc jmfit} were scaled by the above correction
factors to obtain the final flux densities at each frequency.


\tabletypesize{\scriptsize}
\begin{deluxetable*}{lcccc}[t!]
\centering
\tablecaption{Evolution of the spectral index in  V404~Cygni at various epochs
\label{tab:alpha}}
\tablewidth{0pt}
\tablehead{
\colhead{Mean date of} & \colhead{MJD} & \colhead{Days since}   &\colhead{Spectral index} & \colhead{Spectral index\tablenotemark{a}} 
 \\
\colhead{obsn. (UT)} & & \colhead{outburst\tablenotemark{a}} & \colhead{$\alpha_{235/610}$\tablenotemark{b}} &  \colhead{$\alpha_{610/1280}$\tablenotemark{b}} 
}
\startdata
2015 June 26.89 & 57199.89 & 11.12 &  $0.96\pm0.19$ & $0.62\pm0.20$ \\
2015 July 01.93 & 57204.93 & 16.16 &   $-0.43\pm0.22$ & $-0.44\pm0.20$  \\
2015 July 11.02 & 57214.02 & 25.25 &   $\cdots$ &   $> -0.19$ \\
2015 July 12.65 & 57215.65 & 26.88 &  $> -2.07$ & $\cdots$   \\
2015 July 12.93 & 57215.93 & 27.16 &  $\cdots$ &   $0.13\pm0.56$ \\
\enddata
\tablenotetext{b}{Days since the outburst on 2015 June 15.77 UT (MJD 57188.77).}
\tablenotetext{a}{Here $\alpha_{235/610}$ is the spectral index between 235 MHz and 610 MHz and $\alpha_{610/1280}$ is the spectral index between 610 
MHz and 1280 MHz.}
\end{deluxetable*}

The uncertainty in the  flux densities was estimated by measuring the flux densities 
of three random bright point sources in the field of view at different epochs, at the same 
frequency. We find 
that the flux densities of these sources at different epochs match within $\approx 10$\% and 
infer that systematic errors contribute an uncertainty of $\approx 10$\% to our flux density 
estimates. Our final errors on the flux density of \vcyg\ were obtained by summing (in quadrature) 
the measurement error from the single-Gaussian fits and the two sources of systematic error in 
the flux density scale, $\approx 10$\% from the flux density variations between epochs and 
$\approx 3-10$\% from the corrections for the different system temperatures between \vcyg\ 
and the calibrators. Table \ref{tab:obs} lists the details and the results of the various GMRT 
observations.

We searched for intra-epoch variability on time scales of $\approx 15-30$~minutes in the 
two epochs immediately after the outburst of June~25. For the run on June~26, this was done 
by splitting each visibility dataset (at each frequency) into 15-minute chunks, and then carrying 
out the imaging and self-calibration procedure for each 15-minute piece independently. For the 325~MHz 
observing run on June~27, the search for variability was carried out on time scales of 30~minutes. 
The flux densities of \vcyg\ from the different $15-30$~m intervals were found to be consistent 
(within $\approx 1\sigma$ significance) with each other. 
No evidence for intra-observation variability in \vcyg\ was seen in the GMRT data at any epoch.


\section{Results and Discussion}
\label{results}

Our GMRT monitoring of \vcyg\ began on June~26.89~UT, 11.08~days after the first major X-ray outburst of 
June~2015, and within a day of the brightest X-ray flare \citep[on June~25.93~UT; ][]{segreto15,trushkin15a,radhika16}. 
Fig.~\ref{fig:lc} shows the GMRT light curves of \vcyg\ at 235, 610 and 1280~MHz.  The radio emission 
was found to be very bright at the first epoch, followed by a rapid decline at all three frequencies, with the 
decline steepest at 1280~MHz and slowest at 235~MHz. The 1280~MHz flux density decreases by a factor of 
$\approx 100$ within $\approx 4$ days of our first observing epoch, while the 610 and 235~MHz flux 
densities decrease by factors of $\approx 50$ and $\approx 15$, respectively, over the same period. 

\begin{figure*}[t!]
\centering
\includegraphics*[width=0.45\textwidth]{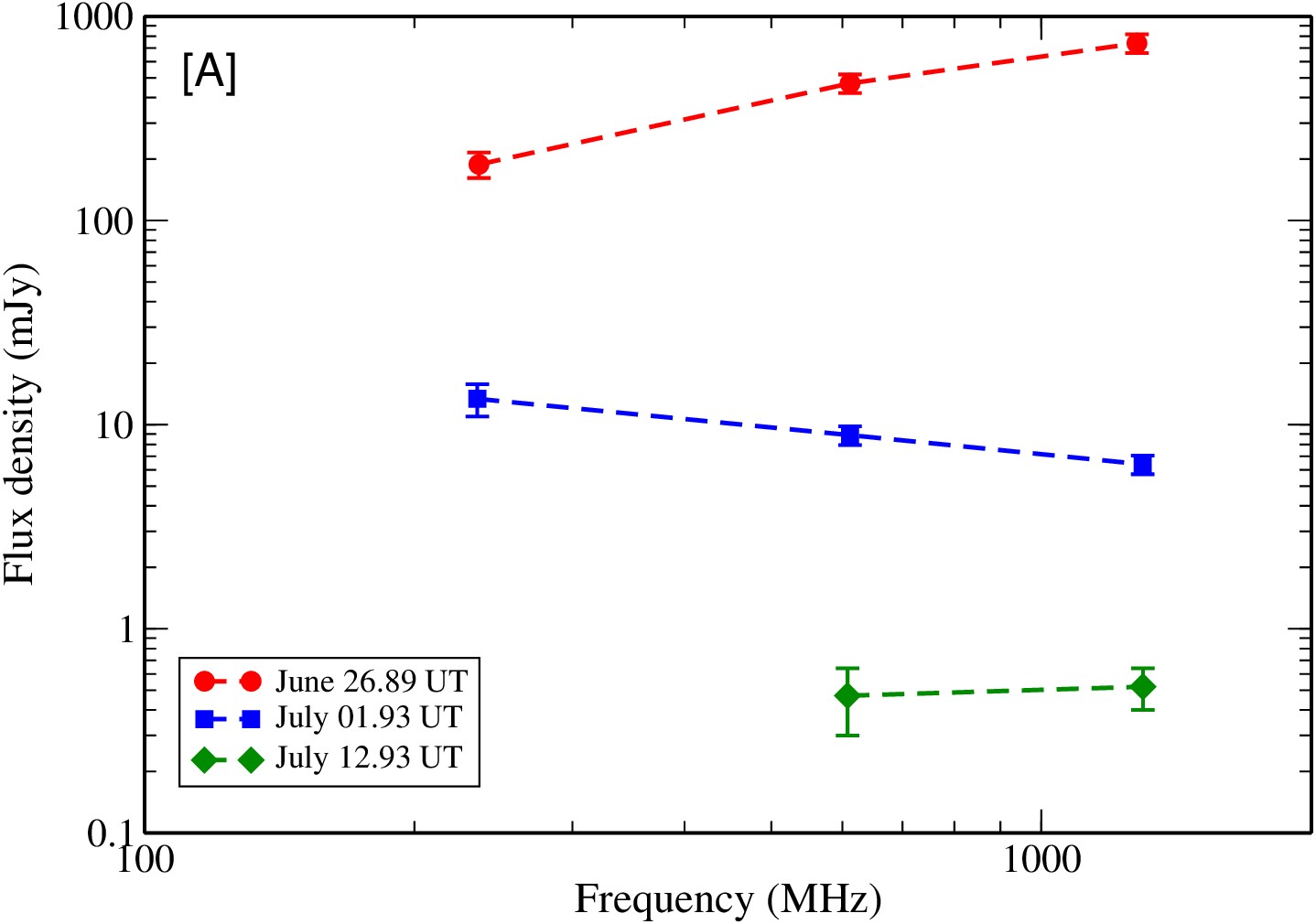}
\hskip 0.1in
\includegraphics*[width=0.45\textwidth]{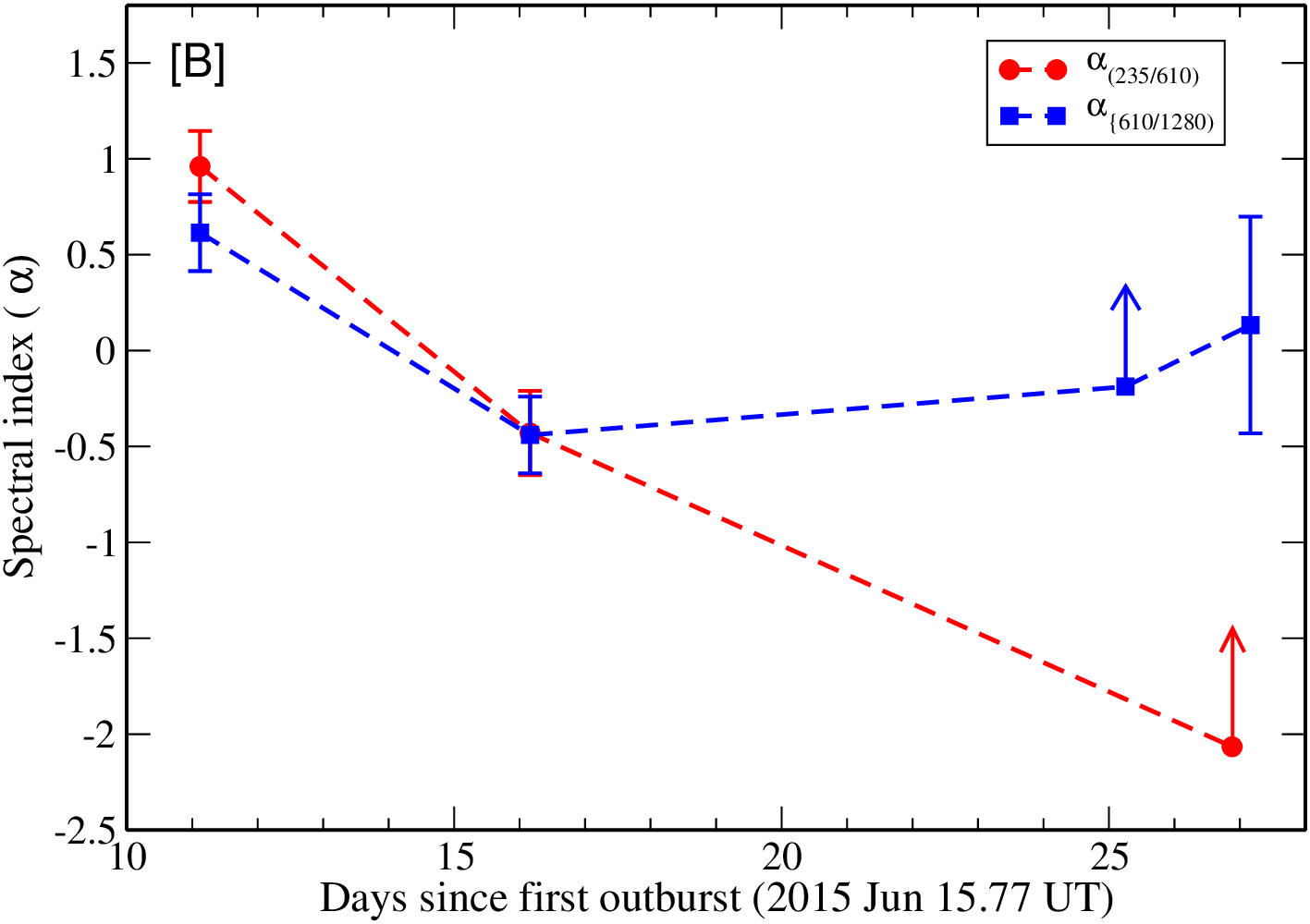}
\caption{[A]~GMRT spectra of V404~Cygni at three different epochs (June~26 in red circles,
July 1 in blue squares, and July 12 in green diamonds), showing the temporal 
evolution of the radio spectrum (using data with two or more detections from Table~\ref{tab:obs}). 
[B]~Evolution of the spectral index of V404~Cygni, between 235 and 610 MHz (red circles), and 
610 and 1280 MHz (blue squares).
\label{fig:spectrum}}
\end{figure*}

Fig.~\ref{fig:spectrum}[A] plots the spectrum of \vcyg\ at three representative epochs, 
June~26.89 UT, July~01.93 UT, and July~12.93 UT, $\approx 11$, $16$, and $27$~days after the 
original outburst.
A change in the spectral shape is clearly present, in addition to the decline in the flux density 
with time. Fig.~\ref{fig:spectrum}[B]  and Table \ref{tab:alpha} show the temporal evolution of the spectral index $\alpha$ 
(defined by $S_\nu \propto \nu^\alpha$, where $S_\nu$ is the flux density at the frequency $\nu$),
evaluated between 235~MHz and 610~MHz ($\alpha_{(\rm 235/610)}$), and 610~MHz and 1280~MHz 
($\alpha_{(\rm 610/1280)}$).  The spectrum is clearly inverted on June 26.89~UT, with positive
spectral indices, $\alpha_{(\rm 610/1280)}=0.61 \pm 0.20$ and $\alpha_{(\rm 235/610)} =0.96 \pm 0.19$,
indicating that the radio outburst is in the optically-thick regime at the GMRT frequencies. 
Further, the fact that $\alpha_{(\rm 610/1280)} < \alpha_{(\rm 235/610)}$ suggests that the 
spectrum is likely to turn over at frequencies slightly above 1.28~GHz. The spectrum is much flatter 
by July~01.93~UT, with $\alpha_{(\rm 610/1280)}=-0.43\pm0.22$ and $\alpha_{(\rm 235/610)} =-0.44\pm0.20$. 
The spectrum remains flat at later epochs, with a decline in the flux density.

The X-ray monitoring of \vcyg\ during June and July 2015 is discussed in detail by \citet{radhika16}
and \citet{plotkin17}. \citet{radhika16} find that \vcyg\ was in the hard X-ray 
state for $\approx 3$ days after the initial outburst of June~15.77~UT, then in an intermediate 
state for 9 days, before moving into the soft state on around June 27~UT. The bright radio flares 
of June~19, June~22 and June~25 all arise during the intermediate state. \citet{plotkin17} find 
that the X-ray spectrum continued to soften during July, with an X-ray luminosity that decreased 
(albeit non-monotically) with time. The first epoch of GMRT monitoring of June~26.89~UT was thus 
in the intermediate X-ray state, while all 
later GMRT epochs were in the soft state. This is consistent with the observed change in the 
GMRT spectral index, from highly inverted at the first epoch (due to the transient radio jet 
or ejected plasmon being in the optically-thick regime at the low GMRT frequencies) to flat at 
later epochs, when \vcyg\ was in the soft state. 

Fig. \ref{fig:variability} shows the flux densities measured at our three observing frequencies 
over different 15-minute intervals during our $\approx 2.5$-hour observing run of June~26. We note 
that the error bars on each measurement include the systematic errors on the flux scale discussed in 
the previous section. No evidence for intra-epoch variability is apparent in the data, over 
the $\approx 2.5$~hour duration of the GMRT observations.

\begin{figure}[t!]
\includegraphics*[width=0.45\textwidth]{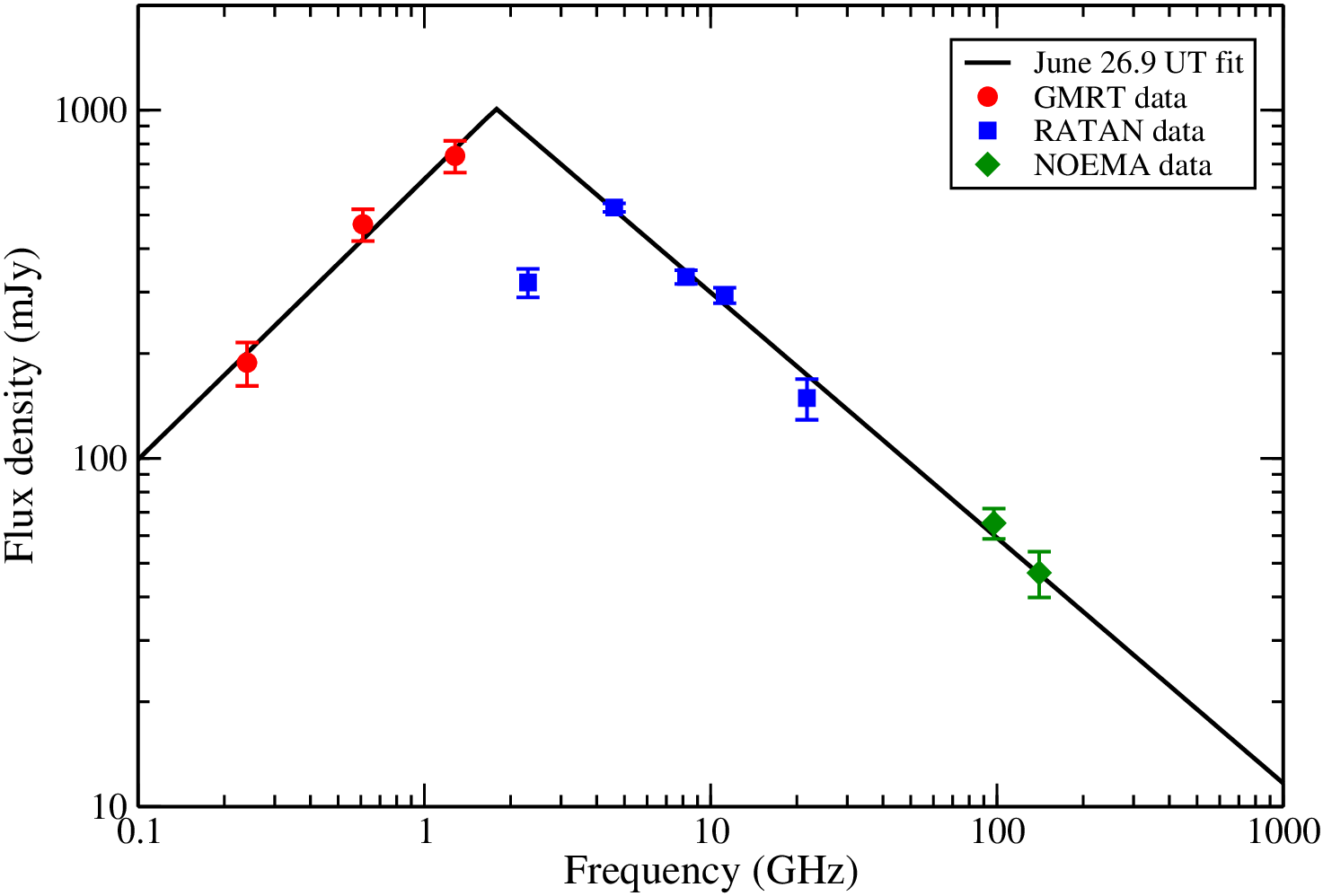}
\caption{The radio spectrum of V404~Cygni on $\approx 2015$~June~26.9~UT. The flux density estimates 
are from the GMRT (red circles, this work), the RATAN-600 dish \citep[blue squares;][]{trushkin15b,trushkin15a} 
and the NOEMA array \citep[green diamonds;][]{tetarenko15}.}
\label{fig:sed}
\end{figure}

Fig.~\ref{fig:sed} shows the spectrum of \vcyg\ obtained from near-simultaneous (within $\approx 0.1$~days) 
measurements between 235~MHz and 140.5~GHz on $\approx $~June~26.9~UT. Besides the low-frequency GMRT 
data (on June~26.89~UT), the figure includes high-frequency flux density estimates from the RATAN-600 telescope 
\citep[at 2.3, 4.6, 8.2, 11.2~ and 21.7~GHz, on June~26.93~UT; ][]{trushkin15b,trushkin15a} and 
from the NOrthern Extended Millimeter Array \citep[NOEMA; at 97.5~GHz and 140.5~GHz, on June 
27.00~UT; ][]{tetarenko15}. It should be emphasized that the different observations are at 
slightly different observing times, within a $\approx 2$~hour period; the GMRT and the RATAN-600 
observations overlap with each other, while the NOEMA observing session took place slightly later.
While no evidence was obtained for variability in the GMRT data over a similar observing period 
(see Fig.~\ref{fig:variability}), we cannot rule out the possibility of variability in the 
high-frequency data on similar timescales. For example, \citet{tetarenko17} find evidence of far more 
short-term variability at high frequencies ($\gtrsim 200$~GHz) than at low frequencies ($\lesssim 30$~GHz), 
during a simultaneous multi-frequency monitoring campaign. The possibility of high-frequency variability 
should be treated as a caveat in the results below. However, the RATAN-600 high-frequency 
observations overlap with the GMRT run, and very similar results are obtained on excluding 
the higher-frequency NOEMA data, which are more offset in time, from our analysis.

The spectrum of Fig.~\ref{fig:sed} shows clear evidence for a spectral turnover below $\approx 1.5$~GHz. 
Interestingly, the RATAN-600 measurements at $\approx 2.3$~GHz and $\approx 4.6$~GHz suggest that 
there may be an additional spectral turnover at $\approx 3$~GHz, followed by a rise in the spectrum 
at $\lesssim 2$~GHz. We emphasize that the evidence for this second turnover is quite tentative, since 
it is effectively based on a single RATAN-600 2.3~GHz data point. If correct, the second turnover 
would suggest that the radio emission arises from two synchrotron self-absorbed regions, perhaps 
due to two separate radio outbursts.

Most attempts to separate between the two standard mechanisms for radio jet launching and collimation, 
the Blandford-Znajek ``spin-powered'' model \citep{blandford77} 
and the Blandford-Payne ``accretion-powered'' model \citep{blandford82} are based on the 
expected correlation between the total jet power and the black hole spin parameter in the 
former class of models. Tentative claims, using a handful of BHXBs, have been made for 
a relation between the transient jet ejection energy or luminosity and the black hole spin
\citep[e.g.][]{fender10,narayan12,steiner13}, but it has also been argued that no single 
relation exists between the transient jet power and the spin parameter \citep{russell13}.

An estimate of the flux density at the turnover frequency between the optically-thick and 
optically-thin regimes can be used, along with assumed equipartition between the particle 
energy and the magnetic field energy, to determine the radius $R$ of the radio jet,
the magnetic field strength $B$, and the total jet power in a radio flare 
\citep[e.g.][]{pacholczyk70,chevalier98,duran13}. Of course, the data of June~26.9~UT 
indicate two synchrotron self-absorbed regions, with two turnover frequencies. However, there 
is only a single flux density measurement in the trough between the two peaks, and the shape 
of the spectrum is not well constrained. We hence chose to fit the spectrum with a 
function of the form

\begin{equation}
f_{\nu}= \begin{cases}
f_{{\rm pk}}\left(\displaystyle \frac{\nu}{\nu_{{\rm pk}}}\right)^{\alpha},
     & \nu < \nu_{{\rm pk}}    \\
f_{{\rm pk}}\left(\displaystyle \frac{\nu}{\nu_{{\rm pk}}}\right)^{-\beta},
     &  \nu > \nu_{{\rm pk}}  \: , 
\end{cases}
\end{equation}

Our best-fit model, with reduced $\chi^2 =0.92$, yields power-law indices $\alpha = 0.81 \pm 0.10$, 
$\beta = 0.71 \pm 0.03$, a turnover frequency of $\nu_{\rm pk} = 1.78 \pm 0.14$~GHz, 
and a flux density $f_{\rm pk} = 1009 \pm 58$~mJy at the turnover frequency. This fit is 
indicated by the solid line in Fig.~\ref{fig:sed}. We note, in passing, that all the flux 
density measurements between 235~MHz and 140~GHz, except for the 2.3~GHz data point, appear 
well-fit by our simple model; this suggests that high-frequency variability is unlikely to 
be a serious issue.

A critical question in the ``equipartition method'' is whether the outflowing plasma 
is moving at relativistic speeds \citep[e.g.][]{duran13}. In the case of BHXBs, 
typical estimates of the bulk Lorentz factor $\Gamma$ lie in the range $\Gamma \approx 1-2$ 
\citep[e.g.][]{hjellming81,hjellming95}. Indeed, for \vcyg, the bulk speeds of multiple 
early outflows in the June~2015 outburst have been found to be low, 
$\Gamma \approx 1-1.3$ \citep{tetarenko17}. We will hence assume that the 
outflowing plasma is moving non-relativistically, using the Newtonian expressions 
\citep{duran13} to estimate $R$, $B$, and the minimum total energy $E_{\rm Eq}$. 
We note the caveat that it is possible that the jet speeds were variable in the ejections
\citep[e.g.][]{tetarenko17}, which would affect the estimates below.

Assuming a self-absorbed synchroton-emitting plasma in a non-relativistic outflow 
with a power-law distribution of electron energies with a spectral index $p > 2$, 
we use equations (16) and (19) of 
\citet{duran13} and the known distance to \vcyg\ \citep[2.39~kpc; ][]{miller-jones09} to estimate 
$R$, $B$, and $E_{\rm Eq}$ from the values of $f_{\rm pk}$ and $\nu_{\rm pk}$. This yields 
$R \approx 4.0 \times 10^{13}$~cm, $B \approx 0.25$~G 
and $E_{\rm Eq} \approx 1.7 \times 10^{39}$~erg. The estimated jet radius during the outburst 
phase is similar to the size of the quiescent radio jet in \vcyg, $\lesssim (4.5-5.0) \times 
10^{13}$~cm \citep{miller-jones08,plotkin17}. Note that assuming $\Gamma \approx 2$ 
or including a second synchrotron self-absorbed component would not significantly increase the 
energy estimate. 

The above estimate of the minimum total energy assumes that all the particle energy is 
in the electrons, which is unlikely to be the case \citep[e.g.][]{duran13}. In the case of 
shock-heated gas, observations suggest that the energy in hot protons is likely to be about 
an order of magnitude larger than that in the electrons  \citep[e.g.][]{panaitescu02,duran13}. 
Assuming that the energy in the protons is ten times larger than that in the electrons 
yields a total minimum energy of $E_{\rm Eq} \approx 7 \times 10^{39}$~erg and an 
equipartition magnetic field of $B \approx 0.5$~G. 

To estimate the minimum jet power in the flare from the minimum total energy $E_{\rm Eq}$, we 
need to know the time $\Delta t$ since the onset of the flare. The brightest X-ray flare 
from \vcyg\ in the June~2015 outburst occurred on June 25.93~UT, approximately a day before 
our first GMRT observations, and this was accompanied by a radio outburst detected with the 
RATAN-600 telescope \citep{trushkin15b,trushkin15a}. We hence assume that $\Delta t \approx 1\:{\rm day}$, 
to obtain a total minimum jet power of ${\rm P}_{\rm jet} \approx 8 \times 10^{34}$~erg~s$^{-1}$, 
again assuming that the energy in hot protons is an order of magnitude larger than that in electrons.

The mass of the black hole in \vcyg\ is $9.0 M_\odot$ \citep{khargharia10}, implying an Eddington 
luminosity of ${\rm L}_{\rm Edd} \approx 1.13 \times 10^{39}$~erg~s$^{-1}$. The ratio of 
transient jet power to Eddington luminosity is thus quite low for the brightest flare of the June~2015 outburst,
Log$[{\rm P}_{\rm jet}/{\rm L}_{\rm Edd}] \approx -4.2$. Similarly low jet powers were obtained 
for \vcyg\ in the June~2015 outburst by \citet{tetarenko17}, based on their simultaneous 
multi-frequency observations of June 22. 

The black hole spin of \vcyg\ has recently been estimated to be $a* > 0.92$ via reflection modelling 
of {\it NuStar} X-ray data \citep{walton17} on
the June~2015 outburst. While \vcyg's spin parameter estimate lies at the upper end of the distribution
of known BHXB spin parameters, it can be seen from Fig.~1 of \citet{russell13} that our estimate of 
the transient jet power for the flare of June~25 is the lowest of the transient jet power estimates. 
Our results for \vcyg\ support earlier studies that suggest that the jet power does not 
correlate with the black hole spin parameter \citep[e.g.][]{fender10,russell13}. However, 
we note that \vcyg\ may well be a unique BHXB, with results for this system non-canonical in 
nature.

The radio emission of \vcyg\ has been found to show rapid intra-day variability in the 
quiescent state \citep[e.g.][]{rana16,plotkin17}. For example, 
\citet{plotkin17} found evidence of strong intra-day variability in the last 
of their VLA monitoring sessions, on August~5, 2015, weak evidence of variability 
(at $\approx 4\sigma$ significance) on August~1, 2015, and no evidence of variability 
on July~28, 2015. We find no evidence of such intra-day variability in our observations of 
June 26 and 27, within two days of the outburst of June~25.93 UT. Interestingly, \citet{tetarenko17}
found much more temporal structure in their high-frequency ($\gtrsim 200$~GHz) data 
than in their low-frequency ($\lesssim 30$~GHz) data, in a simultaneous monitoring 
campaign covering $5-666$~GHz on June 22 UT. The low-frequency light curves appeared
to be smoothed, delayed versions of the high-frequency light curves, consistent with 
a model of multiple expanding jet ejection events \citep{tetarenko17}. Our non-detections
of variability in the low-frequency GMRT data on June 26 UT and later epochs are also 
consistent with such a model.

In summary, we report low-frequency GMRT monitoring of the BHXB \vcyg\ over June~26.89~UT to 
July~12.93~UT, 2015, beginning a day after the strongest X-ray and radio flare in the June~2015 
outburst. The spectrum shows clear evidence of synchrotron self-absorption on $\approx$~June~26.9~UT, 
with two peaks between $\approx 1.5-3$~GHz, suggesting two self-absorbed regions, perhaps arising 
from two outbursts. The low-frequency radio spectrum flattens at later times, as \vcyg\ moves to 
the quiescent state.  Assuming energy equipartition between the particles and the magnetic field, and a 
non-relativistic outflow, we infer a jet radius of $\approx 4.0 \times 10^{13}$~cm, a magnetic field of 
$\approx 0.5$~G, and a total (minimum) jet power of $\approx 8 \times 10^{34}$~erg~s$^{-1}$, assuming 
that the bulk of the particle energy is in hot protons. Our estimate of the transient jet power is 
relatively low, Log$[{\rm P}_{\rm jet}/{\rm L}_{\rm Edd}] \approx -4.2$, despite the high black hole 
spin parameter, supporting earlier results that the radio jet power does not correlate with the 
spin parameter.

\acknowledgments

PC and NK acknowledge support from the Department of Science and Technology via 
SwarnaJayanti Fellowship awards (DST/SJF/PSA-01/2014-15 and DST/SJF/PSA-01/2012-13, 
respectively). We thank Sergei Trushkin for providing us with the RATAN-600 flux densities 
on June 26.93~UT. NK thanks Subhashis Roy for discussions on GMRT system temperature 
calibration. We thank the staff of the GMRT that made these observations possible. 
The GMRT is run by the National Centre for Radio Astrophysics of the Tata Institute 
of Fundamental Research. AIPS is produced and maintained by the National Radio 
Astronomy Observatory, a facility of the National Science Foundation, operated under 
cooperative agreement by Associated Universities, Inc..

{\it Facilities:} \facility{Giant Metrewave Radio Telescope}.
\bibliographystyle{apj}



\end{document}